\documentclass[aip, reprint, amsmath, amssymb]{revtex4-1}

\usepackage{graphicx}
\usepackage{dcolumn}
\usepackage{bm}

\usepackage[utf8]{inputenc}
\usepackage[T1]{fontenc}
\usepackage{mathptmx}
\usepackage{etoolbox}

\begin{document}

\title{Aberration-optimized electro-optic time lens with a tunable aperture}

\author{Sanjay Kapoor}
\email[]{sanjay.kapoor@fuw.edu.pl}
\affiliation{Faculty of Physics, University of Warsaw, Pasteura 5, 02-093 Warszawa, Poland}

\author{Filip So\'snicki}
\email[]{filip.sosnicki@fuw.edu.pl}
\affiliation{Faculty of Physics, University of Warsaw, Pasteura 5, 02-093 Warszawa, Poland}
\affiliation{Paderborn University, Integrated Quantum Optics, Institute for Photonic Quantum Systems (PhoQS), Warburger Straße 100, 33098 Paderborn, Germany }

\author{Micha{\l} Karpi\'nski}
\email[]{michal.karpinski@fuw.edu.pl}
\affiliation{Faculty of Physics, University of Warsaw, Pasteura 5, 02-093 Warszawa, Poland}


\date{\today}

\begin{abstract}
 
Time lenses have been recognized as crucial components for manipulating ultrafast optical pulses in various applications, from ultrafast spectroscopy to interfacing of optical quantum systems. However, the existing analytical model for the electro-optic time lens underutilizes its potential. Here, we introduce a tunable time aperture model for sinusoidal time lenses, enabling precise control over the chirp rate without modifying the device. We derive a closed-form expression for the maximum phase error and demonstrate its dependence on the time aperture. We experimentally validate the model by achieving a 1.6-fold enhanced spectral bandwidth compression of Gaussian pulses compared to the conventional approach. Our framework offers a practical tool for designing efficient temporal optical systems, benefiting applications such as temporal imaging and optical signal processing in both classical and quantum optics, where precise control over spectro-temporal properties is crucial.

\end{abstract}


\maketitle 

\section{Introduction}

The optical space-time duality --- an analogy emerging from the mathematical similarity between the diffraction of a paraxial beam and narrow-band pulse dispersion~\cite{kolner1994} --- allows us to borrow well-established concepts from diffraction optics to the time domain. One particularly important concept derived from this duality is the time lens, which imprints a time-dependent quadratic phase onto an optical pulse, effectively introducing a linear chirp. Temporal optical systems that integrate time lenses with dispersive elements have revolutionized the detection, characterization, and manipulation of ultrafast optical signals~\cite{torres2011, salem2013}. For instance, techniques such as temporal imaging~\cite{kolner1989}, time-to-frequency mapping~\cite{azana2003}, and frequency-to-time mapping are vital tools in modern optical communication~\cite{howe2006}.  More recently, these methods have been extended to quantum optics, enabling advancements in quantum information science~\cite{karpinski2021}. Applications include bandwidth compression of single-photon pulses~\cite{lavoie2013, karpinski2017, filip2020}, characterization of high-dimensional entanglement~\cite{tsang2006, donohue2016}, time-bin superposition detection~\cite{widomski2024}, and frequency shifting~\cite{kapoor2025}. Ideally, applications that require time lenses demand deterministic, large-aperture, high-chirp-rate, and aberration-free designs.

A time lens can be implemented using various linear and nonlinear optical processes. The first traveling-wave electro-optic time lens was demonstrated by Brian Kolner in 1988~\cite{kolner1988}, where the necessary quadratic phase is provided locally by the driving sinusoidal voltage. We will refer to this particular implementation as a sinusoidal time lens. The time aperture (the time window in which the phase is approximately quadratic) of a sinusoidal time lens is limited to about 16\% of the period of the driving signal. Other implementations of the time lens using nonlinear optical processes such as cross-phase modulation~\cite{mouradian2000}, sum (difference) frequency generation~\cite{bennett1994, bennett2000}, and four-wave mixing~\cite{salem2008, foster2008, foster2009} offer larger aperture and higher chirping rate over the sinusoidal time lens. Nonetheless, the electro-optic approach remains popular because of its easy reconfiguration, wavelength-preserving operation, and deterministic nature of the phase imprinting. These properties make it particularly suitable for quantum interface applications~\cite{karpinski2021}, especially with the recent growth in thin-film lithium niobate based integrated photonics~\cite{saravi2021}.

The primary limitations of a sinusoidal time lens are its small time aperture and deviations from an ideal quadratic phase, which introduce temporal aberrations~\cite{bennett2001}. While it is known qualitatively that using longer input pulses with the sinusoidal time lens causes distortions in the output pulse spectrum~\cite{kolner1988}, a quantitative analysis of this effect has been lacking. Although aberration-correcting techniques have been proposed by carefully balancing the higher-order dispersion terms with the high-order time-lens terms~\cite{yurchenko1997, deng2010, li2013, schroder2010}, these methods are experimentally challenging to implement. Alternative techniques involve engineering a more accurate quadratic phase, extending the usable time aperture. One such technique incorporates higher harmonics into the fundamental driving signal to reduce phase errors within a user-defined time window~\cite{howe2004, wu2010, camuniez2010}. However, this method results in high peak voltages, which increases RF power consumption by the modulator. A Fresnel-type time lens, which phase-wraps every integral multiple of $2\pi$~\cite{li2014, filip2018, sosnicki2023}, can mitigate high-voltage requirements but demands complex electronic signals. While these approaches significantly enhance the time aperture, they also require broadband electronics, increasing experimental complexity and cost. In contrast, the single-tone operation of the sinusoidal time lens simplifies implementation and reconfiguration.

In this work, we propose a new model to approximate the chirp rate of a sinusoidal time lens by introducing a tunable aperture parameter—without modifying the time lens itself. We analyze the phase error as a function of the time aperture and present a closed-form expression for the maximum phase error. Then, we perform experiments on the spectral bandwidth compression of Gaussian-shaped pulses with different values of time apertures. We also provide a formula to find an aberration-limited optimal time aperture for a given configuration of a sinusoidal time lens, enabling straightforward use of our approach in various time-lensing schemes such as bandwidth/pulse compression and temporal imaging.

The manuscript is structured as follows: in Sec.~II, we present the theoretical background to the electro-optic sinusoidal time lens and discuss the conventional approximation. We then define a tunable aperture, derive the corresponding chirp rate, and perform the phase error analysis. Sec.~III describes the experimental setup for bandwidth compression and discusses the results. Finally, we conclude in the last section.

\section{Mathematical modeling of time aperture and phase error analysis}
An ideal time lens imprints a time-varying quadratic phase $\theta(t)$,
\begin{equation}
\label{eq:timelensing}
  e^{i\theta(t)} = e^{i\frac{1}{2}Kt^2},
\end{equation}
across an optical pulse, which chirps the pulse linearly with a chirp rate, $K$. The time $t$ is measured in the moving frame of the optical pulse. A simple way to implement a time lens is by using an electro-optic phase modulator (EOPM) driven by a radio frequency (RF) sinusoidal signal. These modulators work on the principle of the Pockels effect, in which the effective refractive index of the electro-optic crystal changes linearly with the applied electric field \cite{saleh}. In a traveling-wave configuration, and in the absence of velocity mismatch, the optical pulse and the driving RF signal co-propagate in the modulator. When the center of the optical pulse is synchronized with a peak or trough of the RF signal, the phase acquired by the pulse can be described by the following expression
\begin{equation}
\label{eq:sinusoidalPhase}
    \theta(t) = \pm A \cos(\omega_m t),
\end{equation}
\noindent
where $\omega_m$ is the angular frequency of the modulating RF signal and $A$ is the phase modulation amplitude. The sign of the phase modulation depends on whether a peak or trough of the RF signal is synchronized with the center of the optical pulse. The modulation amplitude $A$ depends on the amplitude of the applied RF voltage $V_m$,
\begin{equation}
    A = \frac{\pi V_m}{V_\pi},
\end{equation}
where $V_\pi$ is the half-wave voltage of the EOPM. Practically, the maximal modulation amplitude is limited by the breakdown voltage of the modulator and thermal effects.
\begin{figure}
    \centering
    \includegraphics[width=\columnwidth]{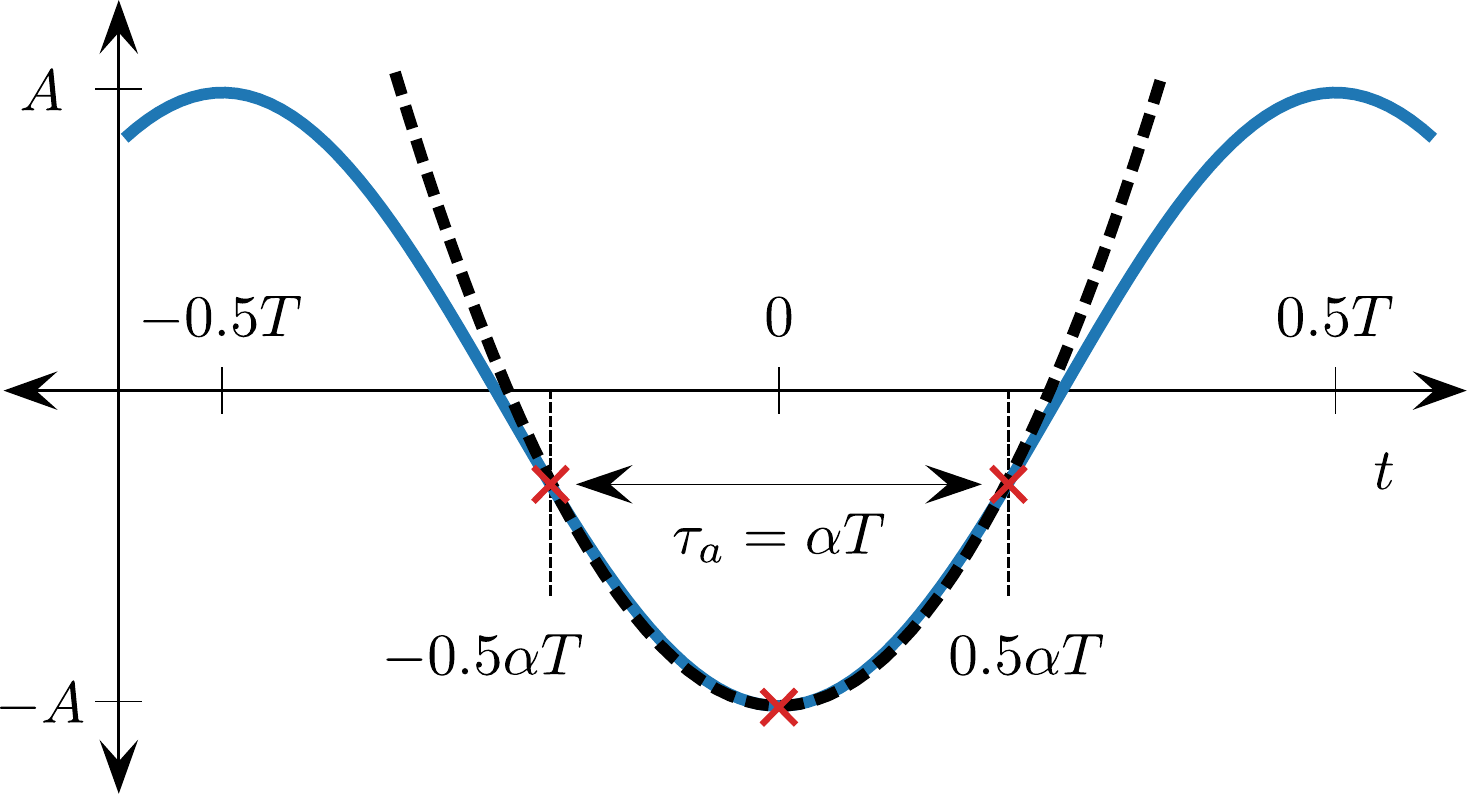}
    \caption{Sinusoidal time lens with a tunable time aperture. The blue solid line depicts the sinusoidal phase, with modulation amplitude $A$ and time period $T$. The two symmetric points (marked by the red crosses) around $t = 0$ define the time aperture ($\tau_a$) as a fraction of the time period ($\alpha T$), $\alpha$ being the aperture parameter. The black dashed line represents the unique quadratic function that passes through the center and both the edges of the time aperture.}
    \label{fig:chirpApprox}
\end{figure}

\subsection{Conventional approximation}
From now on, we will consider the negative sign for the sinusoidal phase modulation such that the resulting chirp rate is positive. When the duration of the optical pulse is much shorter than the period of the RF signal, then the phase modulation can be approximated with the Taylor expansion of the RF signal around a trough,
\begin{equation}
    \theta(t) = - A \left (1 - \frac{\omega_m^2}{2!}t^2 + \frac{\omega_m^4}{4!}t^4 - \cdots \right ).
\end{equation}
We can ignore the first constant term in the expansion because it will introduce only a constant phase shift to the pulse.
Around the cusp, higher-order terms can be ignored, and the phase is approximately quadratic,
\begin{equation}
    \theta(t) \approx  \frac{1}{2}A\omega_m^2t^2,
\end{equation}
implementing a time lens. Comparing Eq.~(5) and  Eq.~\eqref{eq:timelensing} we obtain the chirp rate $K_\text{std} = A\omega_m^2$. Here, we will refer to it as the standard chirp rate. This is the standard formula widely used in the literature \cite{kolner1994, torres2011, salem2013}. There is no precisely defined value for the time aperture of a sinusoidal time lens due to the nature of the approximation. However, within the temporal optics community, it is commonly accepted that a time aperture of $1/\omega_m$ results in a clean output spectrum, which limits the usable aperture to a fixed value of $\approx 16\%$ of the time period of the driving RF signal~\cite{kolner1988, kolner1989, kauffman1993, kolner1994, bennett2001, howe2004, howe2006, pousa2006, li2007, caso2008, camuniez2010, torres2011, salem2013, zhang2013, plansinis2015, mittal2017, zhu2021}.

\subsection{Tunable time aperture for the sinusoidal time lens}

We define the time aperture ($\tau_a$) of a sinusoidal time lens as a fraction of the period ($T = 2\pi/\omega_m$) of the modulating RF signal, parameterized with a dimensionless aperture parameter $\alpha$ as
\begin{equation}
    \tau_a = \alpha T,
\end{equation}
where $\alpha \in (0, 1)$.

We require that the time lens be centered at $t = 0$ and pass through the edges of the time aperture, following the modulating sinusoidal phase, as depicted in Fig.~\ref{fig:chirpApprox}. Next, we derive the equation of the unique quadratic function that intersects these three points, similar to Eq.~\eqref{eq:timelensing}, which provides the chirp rate,
\begin{equation}
\label{eq:newchirp}
    K(\alpha) = A\omega_m^2\text{sinc}^2(\alpha\pi/2).
\end{equation}
In the limit of vanishing aperture parameter
\begin{equation}
   \lim_{\alpha \to 0} K(\alpha) = A\omega_m^2 = K_{\text{std}},
\end{equation}
we obtain the standard chirp rate derived earlier using the Taylor expansion, confirming the consistency of our approach with the standard approximation.

\subsection{Phase error analysis}
\begin{figure}
    \centering
    \includegraphics[width=\columnwidth]{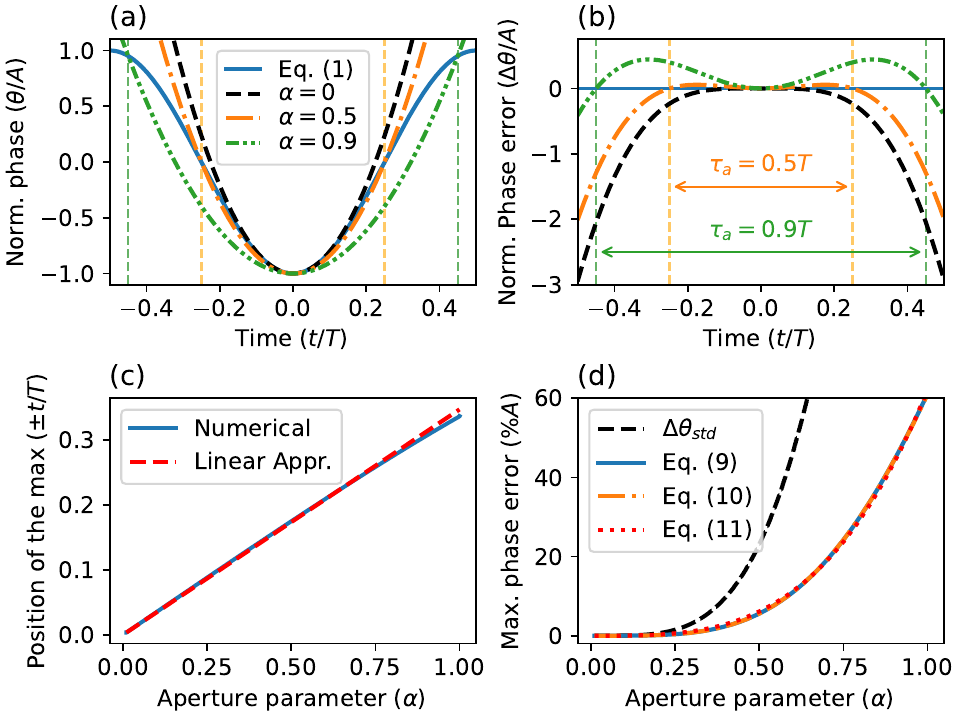}
    \caption{Phase error analysis for different values of the aperture parameter. (a) The sinusoidal phase modulation (solid blue line) and estimated quadratic phases for $\alpha = 0.5$ (orange, dash-dotted line), $\alpha = 0.9$ (green, double-dot-dashed line), and standard formula (black dashed line). The corresponding phase errors are plotted in (b); the thin-dashed vertical lines mark the respective time apertures. (c) The blue solid line is the numerically evaluated position of the maximum phase error as a function of the aperture parameter, and the red dashed line is the linear fit. (d) Comparison of the maximum phase error between the standard chirp rate (black dashed line) and our approximation (blue solid line) as a function of the aperture parameter. The orange dash-dotted line is the approximation for the max phase error described by Eq.~\eqref{eq:maxphase}, and the red dotted line is the power law approximation described in Eq.~(11).}
    \label{fig:phaseResiduals}
\end{figure}

The chirp rate obtained in Eq.~(7) corresponds to an ideal time lens characterized by the chirp rate $K(\alpha)$ and the time aperture $\alpha T$, see Fig.~\ref{fig:phaseResiduals}(a). However, sinusoidal modulation provides a quasi-quadratic phase that has some deviations from the estimated ideal quadratic phase, resulting in aberrations in a temporal optical system. The phase error is the deviation of the sinusoidal phase from a parabolic phase, described as follows:
\begin{equation}
\label{eq:phaseResiduals}
\begin{split}
    \Delta \theta (t/T, \alpha) &= - A\cos(\omega_m t) - \frac{1}{2}K(\alpha)t^2 \\
    & = A \bigg[ -\cos(\frac{2\pi}{T}t) \\
    & - \frac{1}{2} \text{sinc}^2\left(\frac{\alpha \pi}{2}\right) \left(\frac{2\pi}{T}t\right)^2 + 1 \bigg].
\end{split}
\end{equation}
Equation~\eqref{eq:phaseResiduals} shows that the phase error depends linearly on the modulation amplitude. In Fig.~\ref{fig:phaseResiduals}(b), we show the numerically evaluated phase error for $\alpha = 0.5,$ and $0.9$ and compare it with the phase error from the standard approximation. Our approach clearly results in a reduced phase error within the specified time aperture, as indicated by the thin-dashed vertical lines.

For many applications, it is sufficient to consider the peak-to-peak phase error in the desired temporal window. By construction of our approximation, we get zero phase error at the center and edges of the time aperture. We numerically find the position of the maximum phase error (see Fig.~\ref{fig:phaseResiduals}c), which scales linearly with the aperture parameter with a scaling factor $\gamma \approx 0.347$. Thus, the maximum phase error can be evaluated as a function of the aperture parameter by substituting $t/T = \gamma \alpha$ in Eq.~\eqref{eq:phaseResiduals},
\begin{equation}
\label{eq:maxphase}
    \Delta\theta (\alpha, A)_{\text{max}} = 2A[\sin^2(\pi \gamma \alpha) - 4\gamma^2\sin^2(\alpha \pi/2)].
\end{equation}
Alternatively, Eq.~\eqref{eq:phaseResiduals} can be numerically simplified to a power law of the form
\begin{equation}
    \Delta\theta (\alpha, A)_{\text{max}} \approx 0.62 A \alpha^{3.34}.
\end{equation}
We compare the maximum phase error, expressed as a fraction of the modulation amplitude within the defined time aperture, for our approximation and the standard approximation in Fig.~\ref{fig:phaseResiduals}(d). The results show that, for any given modulation amplitude, our approximation allows for a larger aperture for the same level of maximum phase error as that of the standard approach. By extending the usable time aperture, our approach significantly enhances key applications of sinusoidal time lenses, including high-resolution temporal imaging, pulse compression, spectral shaping, and optical signal encoding \cite{torres2011, salem2013}.

The aperture parameter can be selected based on the desired performance and the acceptable level of the maximum phase error. In the next section, we experimentally validate our model for efficient spectral bandwidth compression and provide a formula to balance distortions and performance using analytical and numerical methods.

\section{Spectral bandwidth compression of Gaussian pulses}
\begin{figure*}
    \centering
    \includegraphics[width=0.9\textwidth]{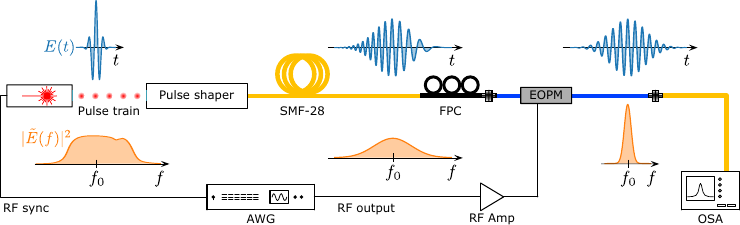}
    \caption{Experimental setup for spectral bandwidth compression. A mode-locked fiber laser generates broadband pulses centered around $1560$~nm with an $80$~MHz repetition rate. A homemade pulse shaper is programmed to produce Gaussian-shaped chirped pulses. A spool of 500-meter-long single-mode fiber (SMF-28) introduces additional group delay dispersion (GDD), further stretching and chirping the pulses. Using a fiber polarization controller (FPC), the polarization of the optical pulses is aligned to the slow-axis of the polarization maintaining fiber of the electro-optic phase modulator (EOPM). The EOPM is driven by a radio frequency (RF) sinusoidal signal generated by an arbitrary waveform generator (AWG) synchronized with the laser, which implements the sinusoidal time lens. The time lens chirps the pulses linearly, resulting in spectral bandwidth compression. The spectrum of the pulses is measured using an optical spectrum analyzer (OSA). When no voltage is applied to the EOPM, it functions as a passive waveguide, allowing for the measurement of the uncompressed spectrum using the same setup. $|\tilde{E}{(f)}|^2$: spectral intensity, $f$: frequency, $f_0$: central frequency, $E(t)$: electric field as function of time. Note that the optical cycles in the pulses are not to scale.}
    \label{fig:setup}
\end{figure*}

The ability to manipulate the spectro-temporal properties of single-photon pulses using phase-only operations is essential for quantum information science \cite{karpinski2021, kielpinski2011}. One crucial application is spectral bandwidth compression, which enables efficient interfacing between spectrally distinct quantum systems \cite{filip2018, sosnicki2023}--such as a parametric down-conversion source and a quantum memory \cite{kimble2008, lvovsky2009}. The concept of bandwidth compression can be understood through the framework of optical space-time duality \cite{torres2011}, where the propagation distance in diffraction optics corresponds to group delay dispersion (GDD) in temporal optics. When a transform-limited (unchirped) pulse travels through a dispersive medium with a GDD of $\Phi$, it expands over time and acquires a linear chirp characterized by a chirp rate of $1/\Phi$. To counteract this effect, a time lens with an equal and opposite chirp rate ($K = 1/\Phi$) is used to unchirp the pulse, which results in compression of the spectral bandwidth. This process is schematically depicted in Fig.~\ref{fig:setup}, which is equivalent to spatial beam collimation using a thin lens, where
\begin{equation}
\label{eq:cc}
    K = 1/\Phi,
\end{equation}
is the temporal collimation condition.

Our objective is to determine the optimal aperture parameter, which is directly related to the optimal GDD through Eq.~\eqref{eq:cc} and Eq.~\eqref{eq:newchirp}). We achieve this by combining analytical and numerical methods with experimental validation. We ultimately provide a general empirical formula to choose the optimal aperture parameter for bandwidth compression using a sinusoidal time lens.

\subsection{Experimental setup}

A detailed schematic of the experimental setup is shown in Fig.~\ref{fig:setup}. The experiments described here were conducted with intense laser pulses, while spectral bandwidth compression principles apply to both intense light pulses and single-photons~\cite{karpinski2017}. The setup consists of three main components: a pulse shaper for preparing Gaussian-shaped pulses, a dispersive medium to introduce GDD, and an electro-optic time lens. 
A mode-locked fiber laser (Menlo Systems, C-fiber HP, $80$~MHz repetition rate) generates broadband pulses of sub-$100$~fs duration centered at $1560$~nm, which are then shaped to Gaussian pulses of varying full-width at half-maximum (FWHM) using a homemade pulse shaper. The pulse shaper is programmed to apply a quadratic spectral phase, producing a tunable GDD in the range of $\approx \pm 6$~ps$^2$. These modulated pulses are then passed through a $500$-m-long SMF-28 fiber to introduce additional GDD of $11.6 \pm 0.4$~ps$^2$, resulting in total tunable GDD in the range of $\approx 5.6$ to $17.6$~ps$^2$.

The chirped pulses are subsequently directed through an EOPM (EOSpace, PM-5SES-20-PFU-PFU-UV-UL) driven by a $16$~GHz sinusoidal signal. The RF signal is generated using an arbitrary waveform generator (Keysight, M8196A) and amplified by $35$~dB using a $3$~W amplifier (Mini-Circuits, ZVE-3W-183+, $18$~GHz bandwidth). This configuration enables a maximum phase modulation amplitude of $\approx 3.7\pi$ radian, as measured by side-band analysis \cite{shi2003}. The modulation frequency is chosen near the amplifier bandwidth to minimize harmonic distortions \cite{baki2006}.

\begin{figure}[t]
    \centering
    \includegraphics[width=0.95\columnwidth]{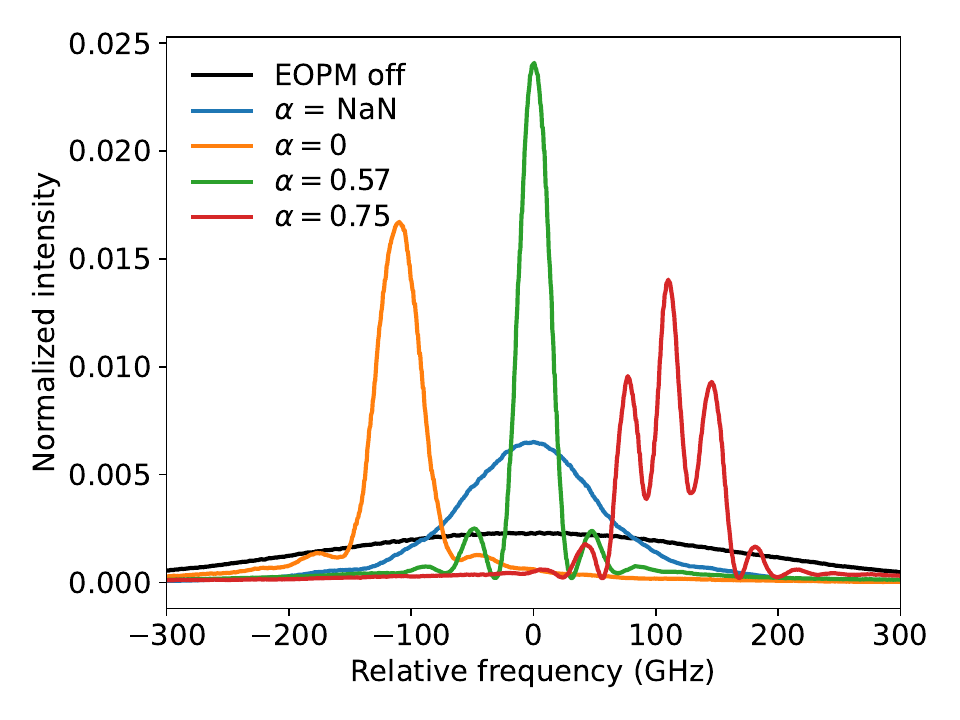}
    \caption{Measured compressed spectra for different aperture parameters for input spectrum of $\approx 418$~GHz ($3.4$~nm) FWHM bandwidth (black dashed trace). Our approximation ($\alpha = 0.57$, green trace) provides better compression than the standard approximation ($\alpha = 0$, orange trace). The trade-off between higher aperture parameter (or higher GDD) and aberrations is noticeable in terms of sidelobes. The spectra in orange and red traces are shifted by $\pm110$~GHz for visual clarity.}
    \label{fig:spectra}
\end{figure} 

\begin{figure*}
    \centering
    \includegraphics[width=0.95\textwidth]{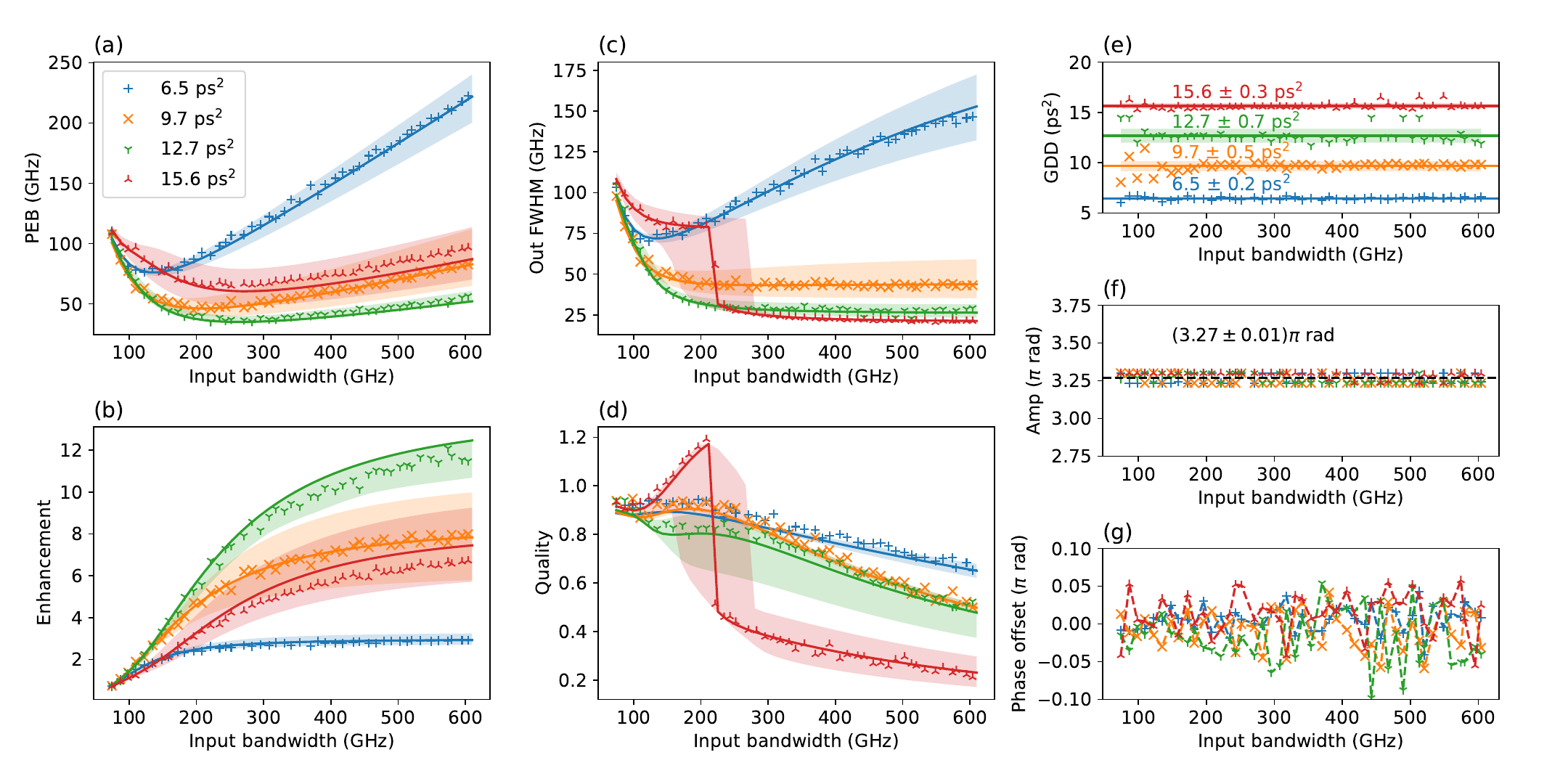}
    \caption{Simulations and experimental results comparing the standard approximation and the new approximation for a sinusoidal time lens with $A = 3.27\pi$ and $\omega_m = 2\pi \times 16$ GHz. (a) Output PEB measured for different values of GDD as a function of FWHM input bandwidth, (b) enhancement,  (c) output FWHM bandwidth, (d) quality of compression, $Q$. Markers represent the measurements, and solid lines are simulations with shaded areas that depict uncertainty obtained from least squares parameter estimation. (e--g) Numerically estimated parameters from the measured spectra using least square minimization.}
    \label{fig:results}
\end{figure*}

The compressed output spectrum is measured using an optical spectrum analyzer (OSA, Yokogawa AQ6370D). The input (uncompressed) spectrum is measured with the RF generator turned off because the EOPM acts as a passive waveguide in the absence of the RF field. Experiments are performed with a fixed $16$~GHz modulation frequency and a phase modulation amplitude of $(3.27 \pm 0.01)\pi$ radians while varying the GDD and input bandwidth. Four GDD values were chosen for the experiment, $9.7$~ps$^2$ (standard GDD, $\alpha = 0$), $12.7$~ps$^2$ (optimal GDD, $\alpha = 0.57$ predetermined numerically), and two extreme values: $6.5$~ps$^2$ ($\alpha$ not defined), and $15.6$~ps$^2$ ($\alpha = 0.75$).

\subsection{Bandwidth compression results and analysis}

The measured compressed spectra for an input Gaussian pulse of FWHM bandwidth $\approx 418$~GHz ($3.4$~nm) are presented in Fig.~\ref{fig:spectra}. The results demonstrate that our approach ($\alpha = 0.57$, green trace) achieves higher spectral compression than the standard approximation ($\alpha = 0$, orange trace). Distortions due to phase errors manifest as symmetric spectral sidelobes. Extending the aperture further ($\alpha = 0.75$, red trace) results in severe spectral degradation, while using a lower dispersion ($\alpha$ undefined, blue trace) yields a cleaner spectrum but with a reduced compression factor.

To quantify spectral bandwidth compression, we introduce the following performance metrics: 

\begin{enumerate}
    \item Power-equivalent bandwidth (PEB):~\cite{saleh}
\begin{equation}
\label{eq:peb}
    B = \frac{\int_{-\infty}^\infty | \tilde{E}{(f)} |^2 df }{\text{max}~| \tilde{E}{(f)}|^2},
\end{equation}
where $|\tilde{E}{(f)}|^2$ is the spectral intensity. In our analysis, we normalize the spectral intensity, i.e., $\int|\tilde{E}(f)|^2df = 1$, so the PEB of a Gaussian spectrum is simply the inverse of its peak spectral intensity. 

\item Relative enhancement ($\eta$):
\begin{equation}
    \eta = \frac{B_\text{in}}{B_\text{out}},
\end{equation}
Where $B_\text{in(out)}$ is the PEB of the input (output) spectrum, the enhancement quantifies the relative gain as a result of spectral compression, enhancing the power spectrum at the central frequency, $f_0$.

\item Quality of compression ($Q$):
\begin{equation}
    Q = \frac{\Delta f_{\text{out}} }{\xi B_{\text{out}}},
\end{equation}
where $\Delta f_\text{out}$ and $B_\text{out}$ are the FWHM bandwidth and PEB of the output spectrum, respectively. $\xi$ is a normalization constant, defined by the ratio of PEB to FWHM, for a Gaussian pulse $\xi \approx 1.06$. $Q = 1$ indicates aberration-free compression, and $Q < 1$ indicates distortions in the output spectrum caused by aberrations.
\end{enumerate}
\par

Figure 5 compares these metrics for different dispersions as a function of input bandwidth. All measured metrics (markers) agree well with the simulations (solid lines), with deviations in the measurements primarily due to thermal fluctuations in the fiber length. The simulations used parameters estimated from the measured spectra through least-squares minimization. The estimated parameters include modulation amplitude, group delay dispersion (GDD), and phase offset, corresponding to the relative delay between the optical pulses and the time lens. These estimated values are displayed in Fig.~5(e--g). The simulations do not account for phase drifts caused by variations in the fiber length due to thermal fluctuations. The shaded regions around the estimated values represent the uncertainty range taken as the standard deviation in the estimated parameters, providing a measure of confidence in the simulations.

In all cases, the relative enhancement asymptotically saturates with the increasing bandwidth of the input spectrum, whereas PEB begins to broaden after reaching a minimum, see Fig.~\ref{fig:results}(a-d). The broadening of PEB indicates the spectral redistribution into the undesired sidelobes due to aberrations in the time lens. Subsequently reducing the quality of compression see Fig.~\ref{fig:results}(d). The quality of compression is best for the smallest GDD, but the enhancement is also low. In contrast, at the intermediate value of $\alpha=0.57$, the enhancement is the highest, and the quality of compression is comparable to the standard approach. We see a $1.6$ times gain in relative enhancement and compression ratio in the FWHM with respect to the standard approach.

In the next section, we describe the procedure to find the input bandwidth and the aperture parameter for any arbitrary value of modulation amplitude.

\subsection{Optimizing the aperture parameter}

\begin{figure}
    \centering
    \includegraphics[width=0.97\columnwidth]{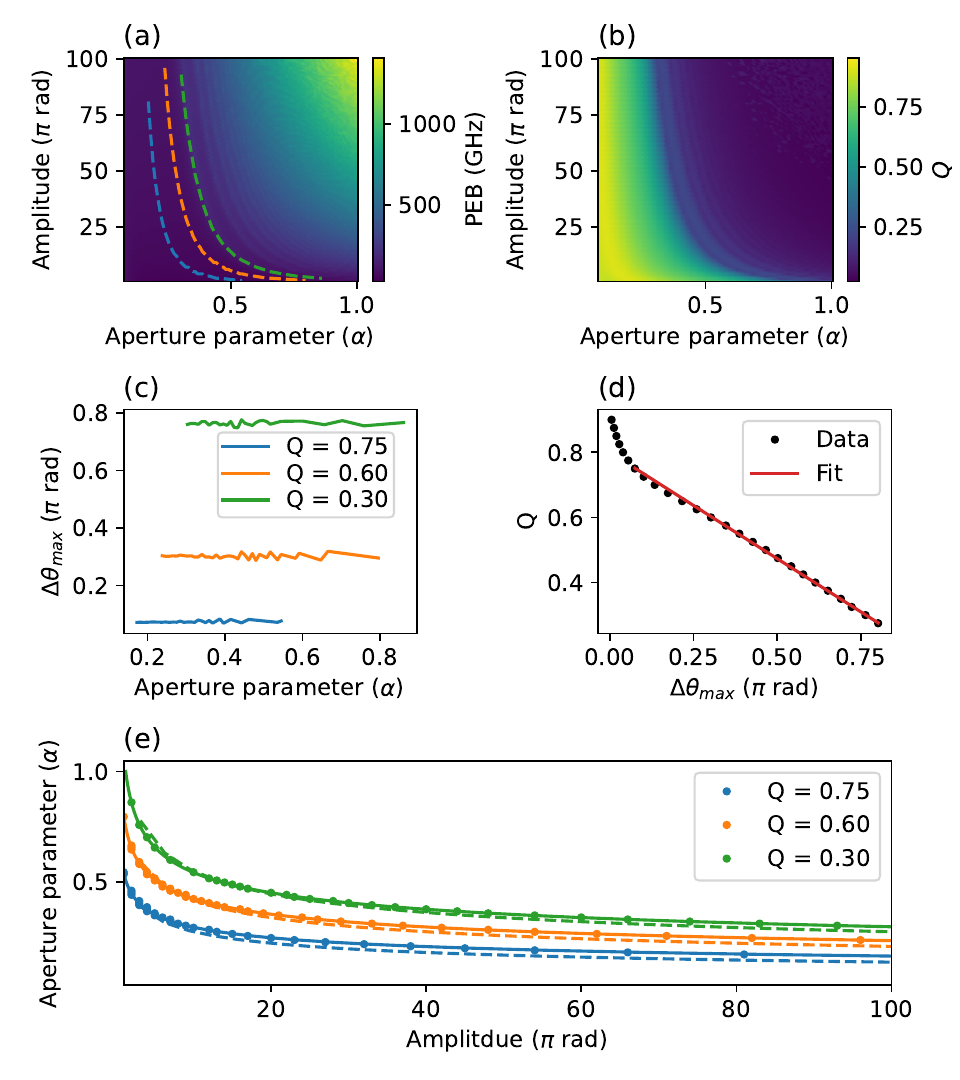}
    \caption{(a) Simulated power-equivalent bandwidth map, and (b) quality of compression map as function of modulation amplitude and the aperture parameter, $\omega_\text{m} = 2\pi \times 16$~GHz, and corresponding $\Delta f_{\text{in}}$ is calculated according to Eq.~(18). The dashed lines in (a) are the constant quality of compression lines for $Q = 0.75, 0.60,$, and $0.3$. (c) Maximum phase error for selected quality of compression. (d) The relationship between maximum phase error and the quality of compression. (e) Relationship between modulation amplitude and aperture parameter for a given quality of compression. The dashed lines are approximated power law for the aperture parameter as a function of modulation amplitude and $Q$; see Eq.~(19).}
    \label{fig:maxAper}
\end{figure}

To establish a relationship between the input bandwidth, $\Delta f_\text{in}$ and aperture parameter $\alpha$, we use the fact that the duration of the chirped pulse must match the aperture of the time lens. A Gaussian pulse with an initial (FWHM) spectral bandwidth $\Delta f_\text{in}$  broadens in time after propagation through a dispersive medium with GDD $\Phi$. In the regime of large GDD, the pulse duration after dispersion is given by~\cite{saleh}
\begin{equation}
\label{eq:pulsedispersion}
    \Delta t_{\text{out}} = 2\pi \Delta f_\text{in} \Phi.
\end{equation}
To ensure optimal spectral compression, we equate the dispersed pulse duration $\Delta t_\text{out}$ with the time aperture of the sinusoidal time lens. Applying the collimation condition, Eq.~\eqref{eq:cc}, and using our chirp rate formula Eq.~\eqref{eq:newchirp}, we can write
\begin{equation}
    \alpha T = \frac{2\pi \Delta f_\text{in}}{A\omega_m^2 \text{sinc}^2(\alpha \pi/2)}.
\end{equation}
Solving for $\Delta f_\text{in}$, we obtain 
\begin{equation}
\label{eq:scalingRule}
    \Delta f_{\text{in}} = A\alpha \omega_m \text{sinc}^2(\alpha \pi/2).
\end{equation}
Equation~\eqref{eq:scalingRule} offers a straightforward method to determine the input bandwidth that maximally utilizes the available aperture, based on the time-lens parameters.

To calculate the optimal aperture parameter $\alpha$ for a given modulation amplitude $A$, we simulate bandwidth compression for a range of aperture parameters and modulation amplitudes. For each value of $\alpha$ and $A$, we calculate the input bandwidth using Eq.~\eqref{eq:scalingRule} for a fixed modulation frequency ($\omega_m = 2\pi\times 16$ GHz) and estimate the required GDD using the collimation condition Eq.~\eqref{eq:cc} and Eq.~(7). 

The results of the simulations are shown in Fig.~\ref{fig:maxAper}. Figure~\ref{fig:maxAper}(a) shows the PEB of the output spectrum as a function of modulation amplitude and aperture parameter. The corresponding quality of compression $Q$ is shown in Fig.~\ref{fig:maxAper}(b). The dashed lines show the constant quality of compression for $Q = 0.75, 0.6,$ and $0.3$. The phase error along these contours, shown in Fig.~\ref{fig:maxAper}(c), suggests that the quality of compression is directly related to the maximum phase error, independent of the modulation amplitude or aperture parameter.

By numerically fitting the relationship between $Q$ and the maximum phase error $\Delta \theta_\text{max}$ (Fig.~\ref{fig:maxAper}d), we find a linear dependence for $Q$ in the range of $0.75$ to $0.3$. Combining this result with the phase error expression in Eq.~(11), we obtain a relationship between modulation amplitude and quality of compression

\begin{equation}
    \label{eq:optimalAlpha}
    \alpha(A, Q) = 1.35 \left( \frac{0.8 - Q}{A}\right)^{0.3}.
\end{equation}
Where $A$ is measured in $\pi$ radians. Equation~\eqref{eq:optimalAlpha} enables us to determine the quality-limited aperture parameter for any value of modulation amplitude. The relationship between $\alpha$ and $A$ for different $Q = 0.75, 0.6,$ and $0.3$ is shown Fig.~\ref{fig:maxAper}(e). The derived expressions offer a practical framework for selecting the aperture parameter that maximizes spectral compression by choosing the quality of compression. This approach provides a systematic way to optimize sinusoidal time lens performance based on the desired balance between bandwidth compression and spectral distortions.

\section{Conclusions and outlook}
In this work, we introduce a new analytical model for estimating the chirp rate of a sinusoidal time lens, incorporating a tunable time aperture. By analyzing the phase error as a function of the time aperture, we derived a closed-form expression for the maximum phase error, enabling the determination of an aberration-limited optimal aperture by balancing the trade-off between time aperture and aberrations. Our approach extends the usable time aperture beyond conventional approximation while maintaining comparable phase error levels, offering significant advantages for sinusoidal time-lens-based applications. We experimentally validate our approach by demonstrating an enhanced spectral bandwidth compression of Gaussian pulses using a sinusoidal time lens. The results show a 1.6 times enhancement in spectral compression compared to the standard approach.

The proposed model has broader implications in classical and quantum temporal optics \cite{karpinski2021, saravi2021, salem2013}, in applications such as temporal imaging \cite{kolner1989, zhu2013, li2015, patera2023}, pulse compression \cite{kolner1988}, time-to-frequency mapping~\cite{azana2003}, and optical signal processing~\cite{howe2006}. Our approach may also be extended to other implementations of time lenses that utilize nonlinear optical methods, such as cross-phase modulation \cite{maram2013, matsuda2016, duadi2021}. Our study provides a practical framework for optimizing temporal systems based electro-optic time lenses. Future research could explore modeling cascaded bandwidth compression setups on integrated photonic circuits for large-scale spectral compression.

\begin{acknowledgments}
This research was funded in part by project QuICHE, supported by the National Science Centre, Poland (NCN, project no.\ 2019/32/Z/ST2/00018), under QuantERA, which has received funding from the European Union’s Horizon 2020 research and innovation programme under grant agreement no.\ 731473, in part by NCN project no.\ 2019/35/N/ST2/04434, in part by the First TEAM programme of the Foundation for Polish Science (project no.\ POIR.04.04.00-00-5E00/18), co-financed by the European Union under the European Regional Development Fund, and in part by the European Union’s Horizon Europe Research and Innovation Programme through grant agreement no.\ 101070700 (MIRAQLS).
\end{acknowledgments}

\section*{Data Availability Statement}

The data that support the findings of this study are openly available in the Dane Badawcze UW data repository of the University of Warsaw at https://doi.org/10.58132/R0CIJX, reference number R0CIJX.


\nocite{*}
\bibliography{references}

\end{document}